\documentclass[%
 reprint,
superscriptaddress,
 amsmath,amssymb,
 aps,
floatfix,
]{revtex4-1}
\usepackage{footnote}
\usepackage{graphicx}
\usepackage{epstopdf}
\usepackage{bm}
\usepackage{mathtools}
\usepackage{textcomp}
\usepackage{amsmath}
\usepackage{physics}
\usepackage{float}
\usepackage{siunitx}
\usepackage{amssymb}
\usepackage{float}
\usepackage{steinmetz}
\usepackage{csquotes}
\usepackage{lipsum}
\usepackage{soul,color}

\newcommand\tab[1][0.8cm]{\hspace*{#1}}

\begin{document}

\preprint{APS/123-QED}
\title{The influence of a planar metal nanoparticle assembly on the optical response of a quantum emitter}

\author{Harini Hapuarachchi}
\email[]{harini.hapuarachchi@rmit.edu.au}

\author{Jared H. Cole}
\email[]{jared.cole@rmit.edu.au}

\affiliation{ARC Centre of Excellence in Exciton Science and Chemical and Quantum Physics, School of Science, 
	RMIT University, Melbourne, 3001, Australia}

\date{\today}

\begin{abstract}
We develop an analytical framework to study the influence of a weakly intercoupled inplane spherical metal nanoparticle (MNP) assembly on a coherently illuminated quantum emitter (QE). We reduce the analytical expressions derived for the aforementioned generic planar setup into simple and concise expressions representing a QE mediated by a symmetric MNP constellation, by exploiting the symmetry. We use the recently introduced generalized nonlocal optical response (GNOR) theory that has successfully explained plasmonic experiments to model the MNPs in our system. Due to the use of GNOR theory, and our analytical approach, the procedure we suggest is extremely computationally efficient. Using the derived model, we analyse the absorption rate, resultant Rabi frequency, effective excitonic energy shift and dephasing rate shift spectra of an exciton bearing QE at the centre of a symmetric MNP setup. We observe that the QE experiences plasmon induced absorption rate spectral linewidth variations that increase in magnitude with decreasing MNP-QE centre separation and increasing number of MNPs. Our results also suggest that, parameter regions where the QE exhibits trends of decreasing linewidth against decreasing MNP-QE centre separation are likely to be associated with plasmon induced excitonic energy redshifts. Similarly, regions where the QE absorption rate linewidth tends to increase against decreasing MNP-QE centre separation are likely to be accompanied by plasmon induced excitonic energy blueshifts. In both these cases, the magnitude of the observed red/blueshift was seen to increase with the number of MNPs in the constellation, due to enhancement of the plasmonic influence.


\end{abstract}

\maketitle
\section{Introduction} \label{Sec:Introduction}
Metamaterials formed by combining different types of nanoparticles are gaining increasing research attention due to their unprecedented capabilities to manipulate light at the nanoscale \cite{hapuarachchi2019analysis, artuso2012optical}. Plasmonic metal nanoparticles (MNPs) \cite{maier2007plasmonics} and quantum emitters (QEs) \cite{premaratne2011light} are two categories of widely studied nanoparticles whose fascinating optoelectronic properties are often expected to synergize when combined \cite{hapuarachchi2018thermoresponsive, senevirathne2019scattering, gettapola2019control, sukharev2017optics}. Due to the tunability of the optical properties using their size and structure, MNPs and QEs possess a wide array of applications in a variety of fields such as biosensing \cite{anker2010biosensing, zhou2015quantum}, photothermal cancer therapy \cite{hapuarachchi2019plasmonic, mallawaarachchi2018superradiant}, optoelectronic nanodevices \cite{jayasekara2015multimode, gamacharige2019significance, kumarapperuma2018complete, weeraddana2016quantum, abeywickrama2019impact} and photovoltaics \cite{tang2011infrared, lin2015electro}. When a QE is kept in nanoscale proximity to a small MNP, a dipole-dipole coupling occurs between the two nanoparticles forming a highly tunable hybrid nanosystem exhibiting interesting optical signatures \cite{hapuarachchi2019analysis, artuso2012optical, sadeghi2010tunable}. It has been shown theoretically, as well as experimentally that hybrid superstructures where one or more MNPs are attached to QEs have the potential to be utilized as versatile sensors and actuators which can surpass the capabilities of the individual constituents \cite{govorov2006exciton, hapuarachchi2019plasmonic}. Therefore, hybrid molecules made of MNPs and QEs have captured the attention of both theorists and experimentalists. Moreover, various techniques to successfully fabricate MNP-QE nanohybrids that can be probed at the single molecule level have already been demonstrated in the literature \cite{hartsfield2015single, zhang2019polarized, nicoli2019dna}. In this context, attempts to enhance our understanding of systems comprising multiple MNPs and QEs are of vital importance \cite{warnakula2019cavity}.

MNPs much smaller than the wavelength of the incident light ($\lambda$) exhibit strong dipolar resonant excitations known as localized surface plasmon resonances (LSPRs) \cite{hapuarachchi2018optoelectronic, liu2017poly}. These resonances enable MNPs to act as nanoscale optical cavities that are able to focus electromagnetic energy to spots much smaller than $\lambda$, overcoming the half-wavelength size limitation of the conventional optical cavities \cite{hapuarachchi2017cavity, ridolfo2010quantum}. LSPRs are nonpropagating modes of excitation of the conduction band electrons which arise naturally from the scattering problem of a subwavelength MNP in an oscillating electromagnetic field. An effective restoring force is exerted on the driven electrons by the curved surface of the particle, which leads to an amplification of the field, both inside the particle, and on the near field of the outside. The resonant condition of this phenomenon is termed a localized surface plasmon resonance \cite{maier2007plasmonics}. In frequency regions close to their plasmonic peaks, MNPs can be used to tailor the optical response of nearly resonant QEs \cite{hapuarachchi2019plasmonic, hapuarachchi2018exciton, hatef2012coherent, hapuarachchi2019plasmonic}. 

It is evident that the optical response of each participating MNP plays a pivotal role in determining the behaviour of the MNP-QE hybrid nanostructures. The most widely adopted analytical approach in the literature to model the optical response of plasmonic nanoparticles such as MNPs is the use of classical local response approximation (LRA) \cite{raza2015nonlocal, govorov2006exciton, artuso2008optical, hatef2012coherent}. However this approach overlooks the nonlocal effects \cite{raza2015nonlocal, mortensen2014generalized} that become prominent in small MNPs, where the ratio of the number of surface atoms to those that make up the bulk of the particle is significant \cite{artuso2012optical}. Thus, LRA has been challenged on a number of accounts, for example its prediction that the surface plasmon resonance energy in the quasistatic limit is independent of the MNP size, which conflicts with the experimentally observed results \cite{raza2015nonlocal, tiggesbaumker1993blue, wubs2017nonlocal, raza2013blueshift, raza2013blueshift2}. Such nonclassical effects could be captured using \emph{ab initio} approaches such as density-functional theory (DFT) \cite{zuloaga2009quantum, andersen2012spatially}. However, such approaches are extremely computationally demanding, especially for nanohybrids formed by coupling several particles together. A simpler and computationally less demanding approach would be to surpass the LRA using nonlocal response theories such as the nonlocal hydrodynamic model or the generalized nonlocal optical response (GNOR) theory \cite{raza2015nonlocal}. The GNOR theory is a recent generalization and an extension of the nonlocal hydrodynamic model, which goes beyond the latter by taking both convection current and electron diffusion phenomena in the MNPs into
account \cite{wubs2017nonlocal, raza2015nonlocal}. It has been shown to better capture both size dependent resonance shifts and linewidth broadening of the MNP extinction cross section
that occurs with decreasing particle size. GNOR also successfully approximates experimentally measured spectra for both monomers and dimers (with nanometer sized gaps) which previously seemed to require microscopic theory and invocation of the quantum mechanical effects of tunneling \cite{raza2015nonlocal, mortensen2014generalized}. 

\begin{figure}[t!]
	\includegraphics[width=\columnwidth]{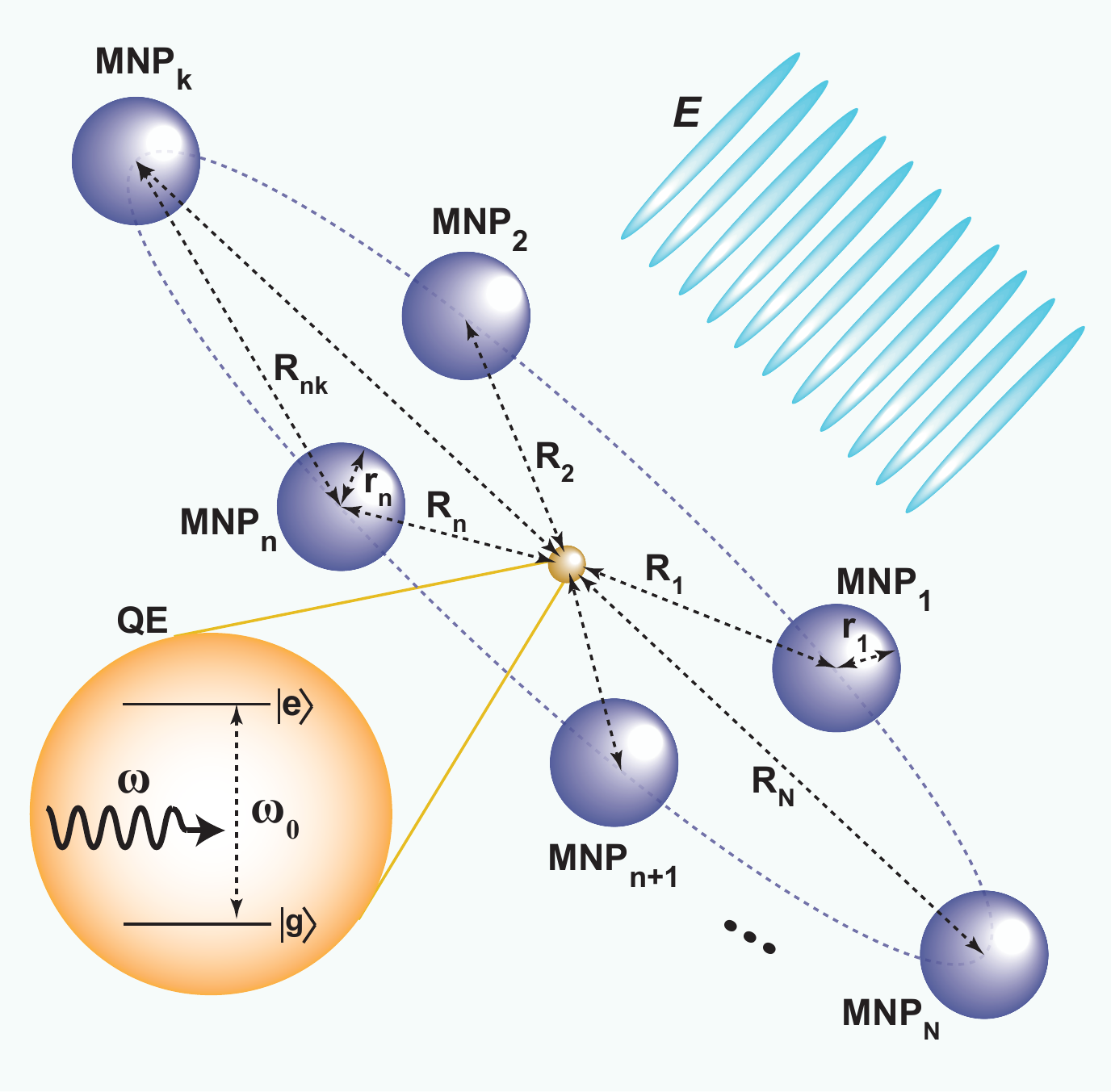}
	\centering
	\caption{(Colour online) Schematic diagram of the system under study. The exciton bearing quantum emitter undergoes dipole interactions with the coherent external drive $\bm{E}$ (incident along the $z$ axis) and the dipole response fields of the metal nanoparticles placed around it, in the $xy$ plane. \label{Fig:Schematic}}
\end{figure}

In this paper, we develop a comprehensive analytical framework to study the interaction of a coherently illuminated two level quantum emitter with a nonlocally modelled inplane assembly of spherical MNPs. We utilize the GNOR theory \cite{mortensen2014generalized, raza2015nonlocal} to model the MNPs in our assembly. The resulting procedure is extremely computationally efficient.

This paper is organized as follows: In section \ref{Sec:Nonlocal_model_for_plasmonic_dipoles}, we first outline the theoretical foundations of modelling MNP dipoles using the GNOR theory, followed by sections \ref{Sec:External_field_induced_plasmonic_dipoles} and \ref{Sec:Quantum_emitter_induced_plasmonic_dipoles}, where we analytically derive the complete forms of the external field induced and QE induced dipoles in the generic planar MNP constellation. In section \ref{Sec:Effective_field_experienced_by_the_QE_exciton}, we derive the effective external field experienced by the coherently illuminated QE, under the influence of the inplane MNP dipole assembly. In section \ref{Sec:Symmetric_Constellation}, we reduce the equations derived in the earlier sections to obtain simplified, elegant analytical equations for a planar symmetric setup. Then we model the QE as an open quantum system in section \ref{Sec:Open_Quantum_System}, and outline the procedure of obtaining the steady state solutions to the QE density matrix in section \ref{Sec:Steady_State_Analysis}. Finally, in section \ref{Sec:Results_and_Discussion}, we use the derived model for a detailed analysis of a QE under the influence of a weakly inter-coupled inplane symmetric MNP setup.

\section{Formalism} \label{Sec:Formalism}
We consider $N$ spherical MNPs of radius $r_\text{n}$, placed at centre-to-centre separations $R_\text{n}$ from a QE, where $\text{n} = {1,...,N}$. This setup, depicted in Fig.\ \ref{Fig:Schematic}, is lying in the $xy$ plane. The exciton bearing quantum emitter (QE) is assumed to have a relatively negligible radius, a relative permittivity $\epsilon_\text{s}$ and an excitonic energy $\hbar\omega_0$. A perpendicularly incident coherent electric field  $\bm{E} = E_0(e^{-i\omega t}+e^{i\omega t})\hat{\bm{\mathit{z}}} = E\hat{\bm{\mathit{z}}}$ optically couples the components of our hybrid nanosystem, where $\omega$ is the angular frequency of the incoming radiation and $\hat{\bm{\mathit{z}}}$ is the unit vector along $z$ axis. The system is submerged in a bath of relative permittivity $\epsilon_\text{b}$ and transient thermal effects \cite{mallawaarachchi2017tuneable, mallawaarachchi2018superradiant} are assumed to be negligible. We assume reasonably sufficient MNP separations to neglect inter-MNP quantum mechanical \cite{zhu2016quantum, wang2020modified} and multipolar hybridization effects \cite{nordlander2004plasmon, su2003interparticle, warnakula2019cavity}, under perpendicular polarization conditions.

Note that throughout the formalism section we use bold fonts, hat notation, tilde notation, plus sign superscript and bold hat notation to refer to vectors, quantum mechanical operators, slowly varying amplitudes, positive frequency components and unit vectors, respectively.

\subsection{Nonlocal model for plasmonic dipoles}\label{Sec:Nonlocal_model_for_plasmonic_dipoles}
For our planar MNP constellation, we consider small metallic spheres, each with radius $r_\text{n}\ll \lambda$ (wavelength of incoming radiation). For such MNPs the ideal dipole representation is valid in the quasistatic regime, which allows for time varying fields but neglects the effects of spatial retardation over the particle volume \cite{maier2007plasmonics}. The system of metal nanoparticles is described by the coupled dipole approximation under the quasistatic limit, where one MNP's influence on the other can be simplified to that of a point dipole at the origin of the former MNP \cite{warnakula2019cavity}. The dielectric permittivity of each MNP is obtained using the Drude-like dielectric function \cite{raza2015nonlocal},
\begin{equation}\label{Eq:Drude}
\epsilon_\text{m}(\omega)  = \epsilon_{\text{core}}(\omega) - \frac{\omega_\text{p}^2}{{\omega(\omega+i \gamma)}},
\end{equation}
where $\omega_\text{p}$ is the bulk plasmon frequency, $\gamma$ is the relaxation constant of the bulk material and $\epsilon_{\text{core}}(\omega)$ is the response of the bound electrons.

Under plane wave illumination with a positive frequency component of the form $\bm{E}_\text{in}^+ = \tilde{\bm{E}}^+_\text{in}e^{-i\omega t}$, an oscillating dipole moment $\bm{d}_\text{n}$ with a positive frequency component of the form \cite{maier2007plasmonics},
\begin{equation}\label{Eq:Induced_dipole}
	\bm{d}^+_\text{n} = (4\pi\epsilon_0\epsilon_\text{b}) r_\text{n}^3\beta_\text{n}\tilde{\bm{E}}^+_\text{in}e^{-i\omega t}
\end{equation}
is induced in the $\text{n}^\text{th}$ MNP, where $\epsilon_0$ is the permittivity of free space and $\beta_\text{n}$ is the \emph{Clausius Mossotti} factor of the $\text{n}^\text{th}$ MNP in a bath of relative permittivity $\epsilon_\text{b}$. The dipole moment $\bm{d}_\text{n}$, the response field $\bm{E}^\text{res}_\text{n}$ of which closely resembles that of a point dipole, can be retrieved as a combination of the positive and negative frequency components as $\bm{d}_\text{n} = \bm{d}^+_\text{n} + \bm{d}^-_\text{n}$, where $\bm{d}^-_\text{n} = (\bm{d}^+_\text{n})^*$ \cite{maier2007plasmonics}. 

The dipole response field $\bm{E}^\text{res}_\text{n}$ emanated by the point-dipole $\bm{d}_\text{n}$ on the centre of another MNP with radius $r_\text{j}$, situated at a radial distance $R_\text{nj}$ can be obtained as \cite{griffiths1999introduction}, 

\begin{equation}\label{Eq:Point_dipole_field}
\bm{E}^\text{res}_\text{n} = \frac{1}{(4\pi\epsilon_0\epsilon_\text{b}) R_\text{nj}^3}\left[3\left(\bm{d}_\text{n}.\hat{\bm{s}}\right)\hat{\bm{s}} - \bm{d}_\text{n}\right],
\end{equation}
where $\hat{\bm{s}}$ is the radial unit vector along the direction of the $\text{j}^\text{th}$ MNP from the centre of the $\text{n}^\text{th}$ MNP. When $\bm{d}_\text{n}$ is perpendicular (parallel) to the axis joining centres of the two MNPs, the above expression simplifies to,
\begin{equation}\label{Eq:Point_dipole_simplified}
\bm{E}^\text{res}_\text{n} = \frac{s_\alpha\bm{d}_\text{n}}{(4\pi\epsilon_0\epsilon_\text{b}) R_\text{nj}^3},
\end{equation}
where the orientation parameter $s_\alpha = -1(2)$ for perpendicular (parallel) polarization. 

The external field feedback dipole induced in the $\text{j}^\text{th}$ MNP due to the above dipole response field $\bm{E}^\text{res}_\text{n}$ of the $\text{n}^\text{th}$ MNP can be obtained by using the positive frequency component of (\ref{Eq:Point_dipole_simplified}) in (\ref{Eq:Induced_dipole}) as,

\begin{equation}\label{Eq:Dipole_induced_dipole}
	\bm{d}_{\text{j}\_\text{n}}^+ = \frac{(r_\text{j}^3\beta_\text{j}s_\alpha)\bm{d}_\text{n}^+}{R^3_\text{nj}}.
\end{equation}

In this study, we use the generalized nonlocal optical response (GNOR) theory \cite{raza2015nonlocal, mortensen2014generalized} to model the \emph{Clausius Mossotti} factor of an MNP with radius $r_\text{n}$ as,
\begin{equation}\label{Eq:NL_Clausius_Mossotti_factor}
\beta_\text{n} = \frac{\epsilon_\text{m}(\omega) - \epsilon_\text{b}(1+\delta_{\text{NL}})}{\epsilon_\text{m}(\omega)+2\epsilon_\text{b}(1+\delta_{\text{NL}})}.
\end{equation}
where the nonlocal correction $\delta_{\text{NL}}$ is given by,
\begin{equation}\label{Eq:delta_NL}
\delta_{\text{NL}} = \frac{\epsilon_\text{m}(\omega) - \epsilon_{\text{core}}(\omega)}{\epsilon_{\text{core}}(\omega)}\frac{j_1(K_\text{L} r_\text{n})}{K_\text{L} r_\text{n} j_1' (K_\text{L} r_\text{n})}.
\end{equation}
In the above equation, $j_1$ denotes the spherical Bessel function of the first kind of angular momentum order 1, $j_1'$ denotes its first order differential with respect to the argument and the longitudinal wave vector abides by the relationship, $K_\text{L}^2 = \epsilon_\text{m}(\omega)/\xi^2\left(\omega\right)$. The bound electron response of the the MNP is obtained as, $\epsilon_{\text{core}} = \epsilon_{\text{exp}}(\omega) + \omega_\text{p}^2\big/\left[\omega(\omega+i\gamma)\right]$ whereas the nonlocal parameter of the GNOR model is characterized by, 
\begin{equation}
	\xi^2(\omega) = \frac{\epsilon_{\text{core}}(\omega) \left[\kappa^2 + \mathrm{D}\left(\gamma - i\omega\right)\right]}{\omega\left(\omega+i\gamma\right)},
\end{equation}
where D is the electron diffusion constant and $\kappa^2 = \left(3\big/5\right)v_F^2$ for $\omega \gg \gamma$ (in the high frequency limit) where $v_F$ is the Fermi velocity of the MNP. It is evident that the \emph{Clausius Mossotti} factor in the conventional local response approximation (LRA), given by the equation $\beta_\text{LRA} = \left[\epsilon_\text{m}(\omega) - \epsilon_\text{b}\right]\big/\left[\epsilon_\text{m}(\omega)+2\epsilon_\text{b}\right]$ \cite{maier2007plasmonics}, can be obtained by setting $\delta_\text{\tiny{NL}}\to 0$ in (\ref{Eq:NL_Clausius_Mossotti_factor}).

\subsection{External field induced plasmonic dipoles} \label{Sec:External_field_induced_plasmonic_dipoles}
We now proceed to compute the effective external field experienced by the QE exciton in the presence of the MNP constellation. Under perpendicular illumination of the external field, all induced electric fields and dipole moment vectors lie parallel to the $z$ axis in the $xy$ plane. Thus, we will only be concerned with their values in the direction of $\hat{\bm{z}}$.

Let us first consider the dipole moment component $d_\text{n}$ directly induced in the $\text{n}^\text{th}$ MNP in the assembly, due to the perpendicularly incident external field. This can be obtained using (\ref{Eq:Induced_dipole}) as,
\begin{equation}\label{Eq:dipole_due_to_external_field}
	d_\text{n} = (4\pi\epsilon_0\epsilon_\text{b}) r_\text{n}^3
\beta_\text{n} E_0 e^{-i\omega t} + c.c., \text{ for } \text{n} = 1,2,...,N
\end{equation} 
where $c.c.$ denotes the complex conjugate of the preceding expression. As the total dipole moment induced in the $\text{n}^\text{th}$ MNP by the external field also comprises feedback via the surrounding MNPs, let us name (\ref{Eq:dipole_due_to_external_field}) as its $0^\text{th}$ level (or direct) external field feedback dipole. 

Using equations (\ref{Eq:Dipole_induced_dipole}) and (\ref{Eq:dipole_due_to_external_field}) as outlined in the appendix, we can find the positive frequency component of the $\text{p}^\text{th}$ level external field feedback dipole formed in the $\text{n}^\text{th}$ MNP due to the collective $(\text{p}-1)^\text{th}$ level external field feedback dipoles in the surrounding MNPs. It takes the following form:

\begin{widetext}
	\begin{equation}\label{Eq:P_th_level_dipole}
	d_{\text{n}\_\Sigma f_{(\text{p}-1)}...\Sigma f_1\_\Sigma f_0}^+ 
	= d_\text{n}^+ s_\alpha^\text{p}\sum_{\substack{f_{\text{p}-1}=1 \\ f_{\text{p}-1} \neq \text{n}}}^N\left(\frac{r_{f_{\text{p}-1}}^3\beta_{f_{\text{p}-1}}}{R_{\text{n}{f_{\text{p}-1}}}^3} \sum_{\substack{f_{\text{p}-2}=1 \\ f_{\text{p}-2} \neq f_{\text{p}-1}}}^N \left(\frac{r_{f_{\text{p}-2}}^3\beta_{f_{\text{p}-2}}}{R_{{f_{\text{p}-1}}{f_{\text{p}-2}}}^3}...\sum_{\substack{f_1=1 \\ f_1 \neq f_2}}^N\left(\frac{r_{f_1}^3\beta_{f_1}}{R_{{f_2}{f_1}}^3}\sum_{\substack{f_0=1 \\ f_0 \neq f_1}}^N \left(\frac{r_{f_0}^3\beta_{f_0}}{R_{{f_1}{f_0}}^3}\right)\right)...\right)\right).
	\end{equation}
\end{widetext} 

The non-truncated form of the total dipole moment induced in the $\text{n}^\text{th}$ MNP due to the external field and its feedback via surrounding MNPs, $d_{\text{n}\_\text{tot}}^E$, can be obtained as,

\begin{align}\label{Eq:d_n_E_tot}
	d_{\text{n}\_\text{tot}}^E &= d_\text{n}^+\left(1+F_{\text{n}\_E}\right) + c.c.,\text{ \tab where, }\\
	F_{\text{n}\_E} &= \frac{\sum_{\text{p}=1}^\infty d_{\text{n}\_\Sigma f_{(\text{p}-1)}...\Sigma f_1\_\Sigma f_0}^+}{d_\text{n}^+}.
\end{align}

For the convergence of dipole feedback in this model, the $\text{p}^\text{th}$ level feedback dipole component arising due to the external field should possess a smaller absolute value than the respective $(\text{p}-1)^\text{th}$ level component. That is,
\begin{equation}\label{Eg:External field_feedback_convergence}
	\abs{d_{\text{n}\_\Sigma f_{(\text{p}-1)}...\Sigma f_1\_\Sigma f_0}^+} < \abs{d_{\text{n}\_\Sigma f_{(\text{p}-2)}...\Sigma f_1\_\Sigma f_0}^+}, \text{  } \forall \text{p}.
\end{equation}

\subsection{Quantum emitter induced plasmonic dipoles}\label{Sec:Quantum_emitter_induced_plasmonic_dipoles}
Due to exciton formation in the QE, a transition dipole moment $d_\text{qe} = \mu\left(\rho_\text{ge}+\rho_\text{eg}\right)$ is assumed to be induced at its centre, where $\mu$ is the off diagonal transition dipole matrix element (assumed real), $\rho_\text{ge} = \tilde{\rho}_\text{ge} e^{i\omega t}$ and $\rho_\text{eg} = \tilde{\rho}_\text{eg} e^{-i\omega t}$ denote the off diagonal density matrix elements of the QE \cite{yariv1967quantum, artuso2012optical}. The positive frequency component of the dipole moment directly induced in the $\text{n}^\text{th}$ MNP due to $d_\text{qe}$ can be obtained with the aid of (\ref{Eq:Dipole_induced_dipole}) as,
\begin{equation}\label{Eq:d_n_qe}
d_\text{n}^{\text{qe}+} = \frac{(s_\alpha r_\text{n}^3\beta_\text{n})\mu \tilde{\rho}_\text{eg}e^{-i\omega t}}{\epsilon_\text{effS}R_\text{n}^3},
\end{equation}
where $\epsilon_\text{effS} = (2\epsilon_\text{b}+\epsilon_\text{s})\big/(3\epsilon_\text{b})$ accounts for the screening of the emanating field due to the QE dielectric \cite{artuso2012optical, hapuarachchi2018exciton,hapuarachchi2019analysis}. Equation (\ref{Eq:d_n_qe}) denotes the 0$^\text{th}$ level feedback dipole formed in the $\text{n}^\text{th}$ MNP due to the QE. 

The positive frequency component of the first level QE feedback dipole formed in the $\text{n}^\text{th}$ MNP due to the collective $0^\text{th}$ level QE feedback dipoles in the surrounding MNPs (indexed using $f_0$ where $f_0\neq \text{n}$) can be obtained using (\ref{Eq:d_n_qe}) and (\ref{Eq:Dipole_induced_dipole}) as,
\begin{equation}\label{Eq:First_level_QE_feedback}
d^{\text{qe}+}_{\text{n}\_\Sigma{f_0}}	= \frac{(s_\alpha^2 r_\text{n}^3\beta_\text{n})\mu \tilde{\rho}_\text{eg}e^{-i\omega t}}{\epsilon_\text{effS}}\sum_{\substack{f_0=1 \\ f_0 \neq \text{n}}}^N\left(\frac{r^3_{f_0}\beta_{f_0}}{R^3_{\text{n}f_0}R^3_{f_0}}\right)
\end{equation}
By repeating this procedure as outlined in the earlier section, we can obtain the $\text{q}^\text{th}$ level QE feedback dipole formed in the $\text{n}^\text{th}$ MNP due to the collective (q-1)$^\text{th}$ level feedback dipoles in the surrounding MNPs (indexed using the subscript $f_{q-1}$ where $f_{q-1}\neq \text{n}$) as,

\begin{widetext}
	\begin{equation}\label{Eq:q_th_level_dipole}
	d_{\text{n}\_\Sigma f_{(\text{q}-1)}...\Sigma f_1\_\Sigma f_0}^{\text{qe}+} 
	= R_\text{n}^3 d_\text{n}^{\text{qe}+} s_\alpha^\text{q}\sum_{\substack{f_{\text{q}-1}=1 \\ f_{\text{q}-1} \neq \text{n}}}^N\left(\frac{r_{f_{\text{q}-1}}^3\beta_{f_{\text{q}-1}}}{R_{\text{n}{f_{\text{q}-1}}}^3} \sum_{\substack{f_{\text{q}-2}=1 \\ f_{\text{q}-2} \neq f_{\text{q}-1}}}^N \left(\frac{r_{f_{\text{q}-2}}^3\beta_{f_{\text{q}-2}}}{R_{{f_{\text{q}-1}}{f_{\text{q}-2}}}^3}...\sum_{\substack{f_1=1 \\ f_1 \neq f_2}}^N\left(\frac{r_{f_1}^3\beta_{f_1}}{R_{{f_2}{f_1}}^3}
	\sum_{\substack{f_0=1 \\ f_0 \neq f_1}}^N \left(\frac{r_{f_0}^3\beta_{f_0}}{R_{{f_1}{f_0}}^3 R^3_{f_0}}\right)\right)...\right)\right).
	\end{equation}
\end{widetext} 

Similar to the external field induced case, the non-truncated form of the total dipole moment induced in the $\text{n}^\text{th}$ MNP due to the QE and its feedback via the surrounding MNPs ($d_{\text{n}\_\text{tot}}^\text{qe}$) can be obtained as,

\begin{align}\label{Eq:d_n_qe_tot}
d_{\text{n}\_\text{tot}}^\text{qe} &= d_\text{n}^{\text{qe}+}\left(1+F_{\text{n}\_\text{qe}}\right) + c.c.,\text{ \tab where, }\\
F_{\text{n}\_\text{qe}} &= \frac{\sum_{\text{q}=1}^\infty d_{\text{n}\_\Sigma f_{(\text{q}-1)}...\Sigma f_1\_\Sigma f_0}^{\text{qe}+}}{d_\text{n}^{\text{qe}+}},
\end{align}
and the QE feedback convergence requires,
\begin{equation}\label{Eg:QE_feedback_convergence}
\abs{d_{\text{n}\_\Sigma f_{(\text{q}-1)}...\Sigma f_1\_\Sigma f_0}^{\text{qe}+}} < \abs{d_{\text{n}\_\Sigma f_{(\text{q}-2)}...\Sigma f_1\_\Sigma f_0}^{\text{qe}+}}, \text{  } \forall \text{q}.
\end{equation}

\subsection{Effective field experienced by the QE exciton}\label{Sec:Effective_field_experienced_by_the_QE_exciton}
The total dipole moment $d_{\text{n}\text{\_tot}}$ induced in the $\text{n}^\text{th}$ MNP in our constellation can be found, while accounting for the infinite series of feedback via the other MNPs, as,
\begin{equation}\label{Eq:total_dipole_in_nth_MNP}
	d_{\text{n}\text{\_tot}} = d^E_{\text{n}\_\text{tot}} + d^\text{qe}_{\text{n}\_\text{tot}}.
\end{equation}

We can then calculate the total electric field incident on the QE exciton, accounting for both the external field $E$ and the collective MNP dipole response fields using (\ref{Eq:total_dipole_in_nth_MNP}) and (\ref{Eq:Point_dipole_field}) as,
\begin{equation}\label{Eq:Tot_field_on_QE_exciton}
E_\text{qe} \approx \frac{1}{\epsilon_\text{effS}}\left( E + \sum_{\text{n}=1}^{N} \frac{s_\alpha d_{\text{n}\text{\_tot}}}{\left(4\pi\epsilon_0\epsilon_\text{b}\right) R_\text{n}^3}\right).
\end{equation}

The resultant Rabi frequency ($\Omega^r$) experienced by the QE exciton  is obtainable using the above equation, where $E_\text{qe} = \tilde{E}^+_\text{qe}e^{-i\omega t} + c.c.$ as,
\begin{equation}\label{Eq:Resultant_Rabi}
	\Omega^r = \mu\tilde{E}^+_\text{qe}\big/\hbar = \Omega + \eta\tilde{\rho}_\text{eg}.
\end{equation}
We can obtain the Rabi frequency in the absence of quantum coherences ($\Omega$) and the QE self-interaction coefficient ($\eta$) using (\ref{Eq:total_dipole_in_nth_MNP}), (\ref{Eq:Tot_field_on_QE_exciton}) and (\ref{Eq:Resultant_Rabi}) as,
\begin{align}
	\Omega &= \Omega^0\left\lbrace 1+\sum_{\text{n}=1}^N\left[\frac{s_\alpha r_\text{n}^3\beta_\text{n}\left(1+F_{\text{n}\_E}\right)}{R_\text{n}^3}\right]\right\rbrace {\text{ and }}\\
	\eta&=\frac{s_\alpha^2\mu^2}{(4\pi\epsilon_0\epsilon_\text{b})\hbar\epsilon_\text{effS}^2}\sum_{\text{n}=1}^N\left[\frac{r_\text{n}^3\beta_\text{n}\left(1+F_{\text{n}\_{\text{qe}}}\right)}{R_\text{n}^6}\right],
\end{align}
where the Rabi frequency experienced by the isolated exciton in the external field $E$ (when $N=0$ or $R\to\infty$) is denoted by $\Omega^0 = \mu E_0\big/(\hbar\epsilon_\text{effS})$ \cite{hapuarachchi2018exciton}. Notice that the normalized Rabi frequency for the case of single MNP-QE nanohybrid \cite{hapuarachchi2018exciton, artuso2012optical} can be retrieved from the above equations by setting $N=1$, where $F_{\text{n}\_E} = F_{\text{n}\_\text{qe}} = 0$. 
 
\subsection{The symmetric constellation}\label{Sec:Symmetric_Constellation}
In the earlier section, we derived the resultant Rabi frequency experienced by a quantum emitter surrounded by a generic planar MNP constellation which was not necessarily symmetric. Let us now reduce the derived equations for a symmetric setup where the QE lies at the centre of a circular constellation of $N$ identical, equidistant MNPs.

Consider a symmetric version of the setup presented in Fig.\ \ref{Fig:Schematic} where $\forall \text{n} \in \{1,2,...,N\}$, radius $r_\text{n}=r$, polarizability $\beta_\text{n}=\beta$, MNP-QE centre separation $R_\text{n}=R$ and the summation of the cube of distances (centre-separations) from the $\text{n}^\text{th}$ MNP to all other MNPs (indexed in the summation as j where j$\neq$n) is commonly given by,
\begin{equation}\label{Eq:zeta}
	\zeta = \sum_{\substack{\text{j}=1 \\ \text{j} \neq \text{n}}}^N \left(\frac{1}{R_\text{nj}^3}\right),
\end{equation}
which will be constant for a given setup due to the circular symmetry. For $N=1$, $\zeta=0$.

Notice that for such symmetric setups, the positive frequency components of the $\text{p}^\text{th}$ level feedback dipole formed in the $\text{n}^\text{th}$ MNP due to the external field, given by equation (\ref{Eq:P_th_level_dipole}), and the $\text{q}^\text{th}$ level feedback dipole formed due to the QE, given by equation (\ref{Eq:q_th_level_dipole}), simplify to their symmetric versions,
\begin{subequations}
	\begin{align}
	\left[d_{\text{n}\_\Sigma f_{(\text{p}-1)}...\Sigma f_1\_\Sigma f_0}^+\right]_\text{sym} &= d_\text{n}^+\left(
	s_\alpha r^3 \beta \zeta \right)^\text{p},\\
	\left[d_{\text{n}\_\Sigma f_{(\text{q}-1)}...\Sigma f_1\_\Sigma f_0}^{\text{qe}+}\right]_\text{sym} &= d_\text{n}^{\text{qe}+}\left(
	s_\alpha r^3 \beta \zeta \right)^\text{q}.
	\end{align}
\end{subequations}
Thus, $F_{\text{n}\_E}$ and  $F_{\text{n}\_\text{qe}}$ reduce to their symmetric versions,
\begin{subequations}
	\begin{align}
	\left[F_{\text{n}\_E}\right]_\text{sym} &= \sum_{\text{p}=1}^{\infty}\left(
	s_\alpha r^3 \beta \zeta \right)^\text{p},\\
	\left[F_{\text{n}\_\text{qe}}\right]_\text{sym} &= \sum_{\text{q}=1}^{\infty}\left(
	s_\alpha r^3 \beta \zeta \right)^\text{q}.
	\end{align}
\end{subequations}
As they both represent infinite complex geometric series, the summations can be simplified to,
\begin{equation}\label{Eq:Geometric_series_sum}
	\left[F_{\text{n}\_E}\right]_\text{sym} = \left[F_{\text{n}\_\text{qe}}\right]_\text{sym} = \frac{s_\alpha r^3 \beta \zeta}{1-s_\alpha r^3 \beta \zeta},
\end{equation}
with the convergence condition,
\begin{equation}\label{Eq:convergence_condition_symm}
\abs{	s_\alpha r^3 \beta \zeta} < 1,
\end{equation}
which determines the usability of this model.

Using (\ref{Eq:total_dipole_in_nth_MNP}) and (\ref{Eq:Geometric_series_sum}), the positive frequency component of the total dipole moment experienced by the $\text{n}^\text{th}$ MNP in a symmetric setup can be obtained as,
\begin{equation}\label{Eq:d_n_tot_symm}
	\left[d_{\text{n}\_\text{tot}}^+\right]_\text{sym} =\frac{d_\text{n}^+ + d_\text{n}^{\text{qe}+}}{1-s_\alpha r^3 \beta\zeta}.
\end{equation}

By substituting $\left[d_{\text{n}\_\text{tot}}^+\right]_\text{sym}$ in place of $d_{\text{n}\_\text{tot}}^+$ in equation (\ref{Eq:Tot_field_on_QE_exciton}), and by setting $R_\text{n}=R$, we can extract the slowly varying amplitude of the total electric field experienced by a QE exciton situated at the centre of our symmetric MNP ring as,
\begin{equation}\label{Eq:Tot_field_on_QE_exciton_symm}
	\left[\tilde{E}^+_\text{qe}\right]_\text{sym} = \frac{\hbar}{\mu}\left[\Omega^r\right]_\text{sym} = \frac{\hbar}{\mu}\left(\left[\Omega\right]_\text{sym} + \left[\eta\right]_\text{sym}\tilde{\rho}_\text{eg} \right),
\end{equation}
where,
\begin{subequations}
	\begin{align}
	\left[\Omega\right]_\text{sym} &=\Omega^0\left\lbrace1 + \frac{N s_\alpha r^3\beta}{R^3}\left(\frac{1}{1-s_\alpha r^3 \beta\zeta}\right)\right\rbrace,\\
	\left[\eta\right]_\text{sym} &= \frac{N s_\alpha^2 r^3 \mu^2 \beta}{\left(4\pi\epsilon_0\epsilon_\text{b}\right)R^6 \hbar \epsilon_\text{effS}^2}\left(\frac{1}{1-s_\alpha r^3 \beta\zeta}\right).
	\end{align}
\end{subequations}

\subsection{The open quantum system}\label{Sec:Open_Quantum_System}
We treat the excitonic system at the centre of the QE quantum mechanically as a two level atom, using the density matrix formalism \cite{blum2012density}. The Hamiltonian of the two level atomic system under the influence of the MNP assembly and the externally applied electric field can be obtained as \cite{weeraddana2017controlling, artuso2008optical, hapuarachchi2018exciton},
\begin{equation}\label{Eq:QE_Hamiltonian}
	\hat{\mathcal{H}} = \hbar\omega_0\hat{\sigma}^+\hat{\sigma}^- - \mu E_\text{qe}(\hat{\sigma}^+ +\hat{\sigma}^-),
\end{equation}
where $\hat{\sigma}^- = |g\rangle \langle e|$, $\hat{\sigma}^+ = |e\rangle \langle g|$, $|g\rangle = (1,0)^\text{T}$ and $|e\rangle = (0,1)^\text{T}$ denote the exciton annihilation and creation operators, and the atomic ground and excited states, respectively. The above Hamiltonian, when taken in isolation describes a closed quantum system where the impact of the environment is not yet taken into consideration. It couples with the environment to form an open quantum system exhibiting irreversible dynamics that can be accounted for using Lindblad terms in the master equation of the QE density matrix $\hat{\rho}$ as follows \cite{hapuarachchi2018exciton, hapuarachchi2017cavity, artuso2012optical},
\begin{equation}\label{Eq:Master_Equation}
	\dot{\hat{\rho}} = \frac{i}{\hbar}\big[\hat{\rho}, \hat{\mathcal{H}}\big] + \lambda_1\mathcal{L}_{\hat{\sigma}^-}+
	\lambda_2\mathcal{L}_{\hat{\sigma}^+}+
	\lambda_3\mathcal{L}_{\hat{\sigma}^+\hat{\sigma}^-}.
\end{equation}
The three Lindblad terms $\lambda_1\mathcal{L}_{\hat{\sigma}^-}$, $\lambda_2\mathcal{L}_{\hat{\sigma}^+}$ and $\lambda_3\mathcal{L}_{\hat{\sigma}^+\hat{\sigma}^-}$ represent bath induced decay of the two level atomic system from the excited to ground state, bath induced excitation vice versa, and elastic scattering processes between the bath and the quantum system, respectively. Their expansion takes the form $\mathcal{L}_{\hat{A}} = 2\hat{A}\hat{\rho}\hat{A}^\dagger - \hat{A}^\dagger\hat{A}\hat{\rho} - \hat{\rho}\hat{A}^\dagger\hat{A}$.

For optical frequencies, even near room temperature, $\lambda_2\approx 0$ \cite{artuso2012optical}. Let, 
\begin{subequations}
\begin{align}
	T_1 &= 1\big/(2\lambda_1),\\
	T_2 &= 1\big/(\lambda_1 + \lambda_3), 
\end{align}
\end{subequations}
where $T_1$ is the energy or population relaxation time of the QE which leads to a mixing between (populations or the diagonal density matrix elements) $\rho_\text{gg}$ and $\rho_\text{ee}$ \cite{artuso2012optical}. $T_2$ is the polarization relaxation or dephasing time which causes losses in the off diagonal density matrix elements \cite{sadeghi2010tunable,kosionis2012nonlocal}. With these definitions, the matrix form of the master equation (\ref{Eq:Master_Equation}) in the basis space $\{ |g\rangle , |e\rangle \}$ reads \cite{hapuarachchi2018exciton, artuso2012optical},
\begin{widetext}
	\begin{equation}\label{Eq:Master_equation_matrix_form}
	\dot{\hat{\rho}} = \frac{i}{\hbar}\begin{bmatrix}
	-\mu E_\text{qe}(\rho_\text{ge}-\rho_\text{eg}) &  -\mu E_\text{qe}(\rho_\text{gg}-\rho_\text{ee})+\hbar\omega_0\rho_\text{ge} \\
	-\mu E_\text{qe}(\rho_\text{ee}-\rho_\text{gg})-\hbar\omega_0\rho_\text{eg} & -\mu E_\text{qe}(\rho_\text{eg}-\rho_\text{ge}) \\
	\end{bmatrix} -
	\begin{bmatrix}
	(\rho_\text{gg}-1)\big/{T_1} & \rho_\text{ge}\big/T_2\\
	\rho_\text{eg}\big/T_2 & {\rho_\text{ee}}\big/{T_1}
	\end{bmatrix}.
	\end{equation}
\end{widetext}
By solving this, we can analyse the behaviour of the two-level excitonic system under the influence of the MNP assembly and the external illumination.

\subsection{Steady state analysis}\label{Sec:Steady_State_Analysis}
In this section, we summarize the approach outlined in \cite{hapuarachchi2018exciton} to obtain the steady state solution of the QE master equation for completeness.

Using element-wise comparison of the left and right hand sides of (\ref{Eq:Master_equation_matrix_form}) and the definition of effective Rabi frequency $\Omega^r=\mu \tilde{E}_\text{qe}^+\big/\hbar$, we can arrive at the following Bloch equations for the two-level excitonic system,
\begin{subequations} \label{Eq:Bloch_equations}
	\begin{align}
	\dot{\rho}_\text{ee} &= -\frac{\rho_\text{ee}}{T_1} + i\Omega^r\rho_\text{ge} - i\Omega^{r*}\rho_\text{eg} \label{Eq:rhoee_tilde_dot},\\
	\dot{\rho}_\text{gg} &= \frac{\rho_\text{ee}}{T_1} - i\Omega^r\rho_\text{ge} + i\Omega^{r*}\rho_\text{eg} \label{Eq:rhogg_tilde_dot},\\
	\dot{\tilde{\rho}}_\text{eg} &=-\left[ i(\omega_0 - \omega) + 1\big/T_2\right]\tilde{\rho}_\text{eg} + i\Omega^r\Delta, \label{Eq:rhoeg_tilde_dot}
	\end{align}
\end{subequations}
where $\Delta = \rho_\text{gg} - \rho_\text{ee}$ denotes the population difference. Defining the real-imaginary separations, $\tilde{\rho}_\text{ge} = \mathcal{A}+i\mathcal{B}$, $\tilde{\rho}_\text{eg} = \mathcal{A}-i\mathcal{B}$, $\Omega = \Omega_\text{re}+i\Omega_\text{im}$ and $\eta = \eta_\text{re}+i\eta_\text{im}$, we can recast (\ref{Eq:Bloch_equations}) to the following form \cite{hapuarachchi2018exciton, artuso2012optical},
\begin{subequations} \label{Eq:Diff_eq_set}
	\begin{align}
	\dot{\mathcal{A}} &= -\frac{\mathcal{A}}{T_2} + \delta \mathcal{B} - \left(\Omega_{\text{im}} + \eta_{\text{im}}\mathcal{A} - \eta_\text{re} \mathcal{B}\right)\Delta \label{Eq:A_dot},\\
	\dot{\mathcal{B}} &= -\frac{\mathcal{B}}{T_2} - \delta \mathcal{A} - \left(\Omega_{\text{re}} + \eta_{\text{re}}\mathcal{A} + \eta_{\text{im}} \mathcal{B}\right)\Delta \label{Eq:B_dot},\\
	\dot{\Delta} &= \frac{1 - \Delta}{T_1} + 4 \left[\Omega_{\text{im}}\mathcal{A} + \Omega_{\text{re}}\mathcal{B} + \eta_{\text{im}}\left(\mathcal{A}^2 + \mathcal{B}^2 \right)\right], \label{Eq:Delta_dot}
	\end{align}
\end{subequations}
where the detuning is denoted by $\delta = \omega-\omega_0$. In the steady state where $\dot{\mathcal{A}} = \dot{\mathcal{B}} = 0$, we can manipulate (\ref{Eq:A_dot}) and (\ref{Eq:B_dot}) to obtain \cite{hapuarachchi2018exciton},
\begin{subequations}\label{Eq:A_and_B}
	\begin{align}
		\mathcal{A} &= -\mathrm{Re}\left(\frac{\Omega\Delta}{\delta + \eta\Delta + i/T_2}\right).\\
		\mathcal{B} &= \mathrm{Im}\left(\frac{\Omega\Delta}{\delta + \eta\Delta + i/T_2}\right).
	\end{align}
\end{subequations}
By substituting (\ref{Eq:A_and_B}) in (\ref{Eq:Delta_dot}) and setting $\dot{\Delta}=0$, we can obtain the following steady state equation for the QE population difference $\Delta$,
\begin{equation}\label{Eq:Detla_polynomial}
	\Delta^3+\bar{w}_2\Delta^2+\bar{w}_1\Delta+\bar{w}_0=0,
\end{equation}
where,
\begin{subequations}
	\begin{align}
		\bar{w}_2 &= \frac{2T_2^2\delta\eta_{\text{re}} + 2T_2\eta_{\text{im}} - T_2^2\left(\eta_{\text{re}}^2 + \eta_{\text{im}}^2\right)}{T_2^2\left(\eta_{\text{re}}^2 + \eta_{\text{im}}^2\right) },\nonumber\\
		\bar{w}_1 &= \frac{T_2(4 T_1|\Omega|^2 - 2\eta_{\text{im}}) + T_2^2(\delta^2 - 2\delta\eta_{\text{re}}) + 1}{T_2^2\left(\eta_{\text{re}}^2 + \eta_{\text{im}}^2\right) }, \nonumber\\
		\bar{w}_0 &= \frac{-T_2^2\delta^2 - 1}{T_2^2\left(\eta_{\text{re}}^2 + \eta_{\text{im}}^2\right) }. \nonumber
	\end{align}
\end{subequations}
By solving (\ref{Eq:Detla_polynomial}) for $-1\leq\Delta\leq 1$ we can obtain the QE population difference, substitution of which in (\ref{Eq:A_and_B}) yields $\mathcal{A}$ and $\mathcal{B}$, and hence the off diagonal density matrix elements $\rho_\text{eg}$ and $\rho_\text{ge}$. 

Let us now look at how the obtained solutions for the density matrix elements can be used to analyse the steady state behaviour of the excitonic system. Firstly, the energy absorption rate of the QE can be computed as \cite{artuso2012optical},
\begin{equation}\label{Eq:QE_Absorption_rate}
	Q_\text{qe} = \hbar\omega_0\rho_{ee}/T_1 = \hbar\omega_0(1-\Delta)/(2 T_1).
\end{equation}
Secondly, we recast the QE Bloch equation (\ref{Eq:rhoeg_tilde_dot}) in the following form,
\begin{equation}\label{Eq:rho_21_dot_recast_2}
\dot{\tilde{\rho}}_\text{eg} = -\left[i(\Pi - \omega) + \Lambda\right]\tilde{\rho}_{eg} + i\Omega\Delta.
\end{equation}
where,
\begin{subequations}
	\begin{align}
	\Pi &= \omega_0 - \eta_{\text{re}}\Delta \label{Eq:Pi_21} \text{\tab and}\\
	\Lambda &= 1/T_2 + \eta_{\text{im}}\Delta \label{Eq:Lambda_21}
	\end{align}
\end{subequations}
denote the effective energy and dephasing rate of the QE excitonic transition under the influence of the neighbouring MNP assembly \cite{hatef2012coherent}. Let us call the two factors $\Pi_f=\eta_{\text{re}}\Delta$ and $\Lambda_f = \eta_{\text{im}}\Delta$ as the exciton transition energy (red) shift and the dephasing rate (blue) shift, respectively \cite{hapuarachchi2018exciton}.

We conclude the formalism section recalling that all the above equations can be converted to their LRA based forms by setting $\delta_\text{NL}\to 0$ as mentioned in section \ref{Sec:Nonlocal_model_for_plasmonic_dipoles}.

\section{Results and discussion}\label{Sec:Results_and_Discussion}

\begin{figure}[t!]
	\includegraphics[width=\columnwidth]{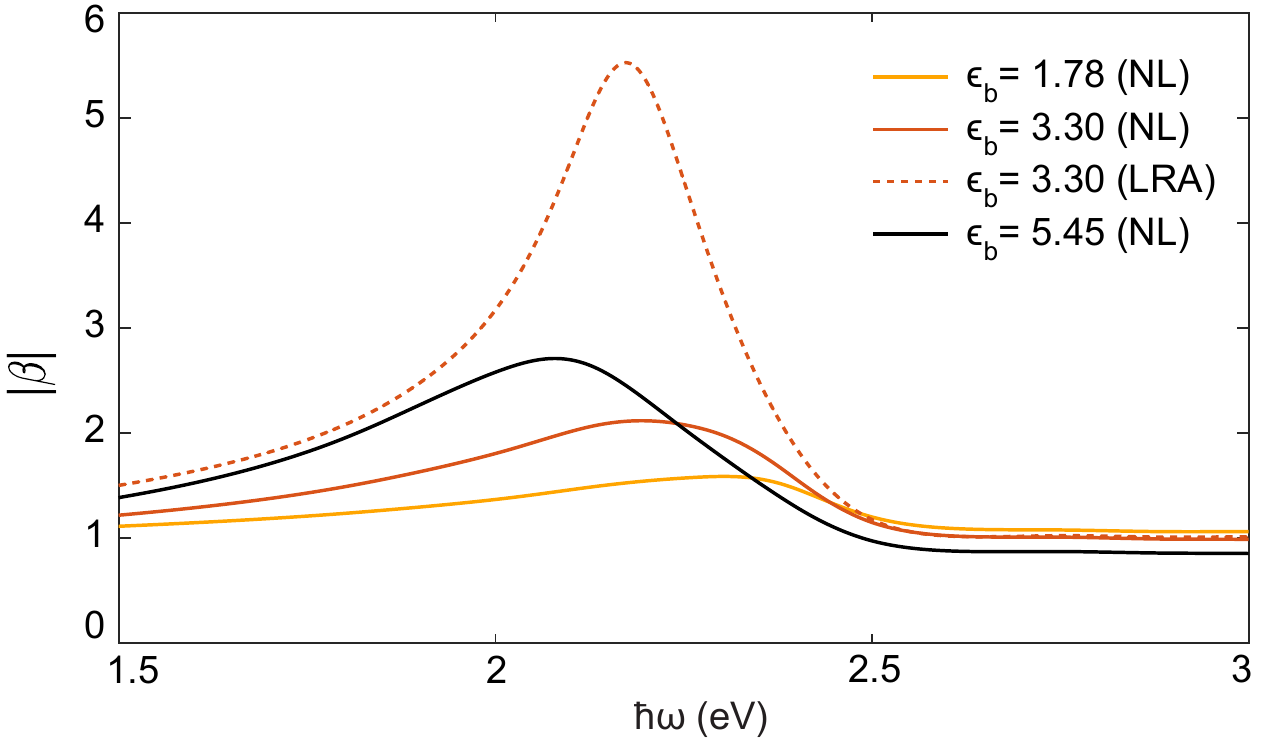}
	\centering
	\caption{(Colour online) Variation of the absolute polarizability ($|\beta|$) for a gold nanoparticle of radius $\SI{3}{\nano\meter}$ as predicted by the nonlocal GNOR (NL) model at submerging medium (relative) permittivities $\epsilon_\text{b} = 1.78$, 3.3, 5.45, and by the local response approximation (LRA) model at $\epsilon_\text{b} = 3.3$.  \label{Fig:beta_comparison}}
\end{figure}

In this section, we study the behaviour of coherently illuminated QEs under the influence of planar, symmetric MNP assemblies using the presented analytical equations. We focus our attention on small MNPs where nonlocal effects are prominent and comparatively large inter-MNP distances within the quasi-static limit where hybridization of the dipole mode of one MNP with higher order multipoles of a neighboring pair (resulting in additional higher order terms to the plasmon coupling \cite{funston2009plasmon}) can be safely neglected. 

The common parameters used for the presented analysis are as follows: electric field strength of the external illumination $E_0 = \SI{1e5}{\volt\per\meter}$, orientation parameter $s_\alpha = -1$ (perpendicular illumination), polarization relaxation (dephasing) time of the isolated QE $T_2 = \SI{0.3}{\nano\second}$, energy or population relaxation time of the isolated QE $T_1 = \SI{0.8}{\nano\second}$ \cite{artuso2012optical}, dielectric constant of the QD material $\epsilon_\text{s} = 6$ and QE dipole moment $\mu=\SI{2}{e.\nm}$ \cite{artuso2012optical}. Our analysis uses an assembly of gold MNPs, each with radius $r=\SI{3}{\nano\meter}$, bulk plasma frequency $\hbar\omega_\text{p} = \SI{9.02}{\electronvolt}$, bulk damping rate $\hbar\gamma = \SI{0.071}{\electronvolt}$ Fermi velocity $v_f = \SI{1.39e6}{\meter\per\second}$ and diffusion constant $\text{D}\approx\SI{8.62e-4}{\meter\squared\per\second}$ \cite{raza2015nonlocal}. The experimentally measured bulk dielectric data $\epsilon_{\text{exp}}$ for gold are obtained using the tabulations by Johnson and Christy \cite{johnson1972optical}.  

In our analysis, we vary the number of MNPs $N$ in our planar symmetric constellation from 1 to 6 to obtain the presented graphical results, while utilizing the symmetry of the hybrid nanosystem to calculate the inter-MNP distances ($R_{\text{n}\text{k}}$), in terms of the centre separation between each MNP and QE ($R$). We list these distances below, in the notations used in our analytical equations, for the convenience of the readers. 
\begin{subequations}
	\begin{align*}
	&\text{For all }\text{n},\text{k}, \text{ }R_{\text{n}\text{n}}=0 \text{ and }R_{\text{n}\text{k}}=R_{\text{k}\text{n}} \\
	&N=2 \Rightarrow  R_{12} = 2R  \\
	&N=3 \Rightarrow R_{12} = R_{13} = R_{23} = \sqrt{3}R\\
	&N=4 \Rightarrow R_{12} = R_{14} = R_{23} =R_{34} = \sqrt{2}R,\\
	&\text{\tab\tab}R_{13}=R_{24}=2R\\
	&N=5 \Rightarrow R_{12} = R_{23} =R_{34} =R_{45}= R_{15}=  2R\sin{\frac{\pi}{5}},\\
	&\text{\tab\tab}R_{13}=R_{14}=R_{24}=R_{25}=R_{35}=2R_{12}\sin{\frac{3\pi}{10}}\\
	&N=6 \Rightarrow R_{12} = R_{23} =R_{34} =R_{45}= R_{56} =R_{16} = R,\\
	&\text{\tab\tab}R_{13}=R_{24}=R_{35}=R_{46}=R_{15}=R_{26}=\sqrt{3}R,\\
	&\text{\tab\tab}R_{14}=R_{25}=R_{36}=2R.
	\end{align*}
\end{subequations}
Thus, It is evident that we can obtain the inter-MNP distances for one side of a symmetric setup as $2 R \sin\left(k\pi/N\right)$ where $k=1,2...,\lfloor N/2\rfloor$ up to and beyond $N=6$. If one intends to use $N > 6$, it is worth noticing that the centre separation between adjacent MNPs will be less than $R$ for a symmetric setup. The distances for such assemblies should be cautiously chosen when using this model to stay within the quasi-static limit while maintaining weak inter-MNP coupling. A surface-separation/diameter ratio at least $\geq 1$ between adjacent gold MNPs is consistent with weak near-field plasmon coupling \cite{su2003interparticle} where our model is applicable.   

\begin{figure*}[t!]
	\includegraphics[width=\textwidth]{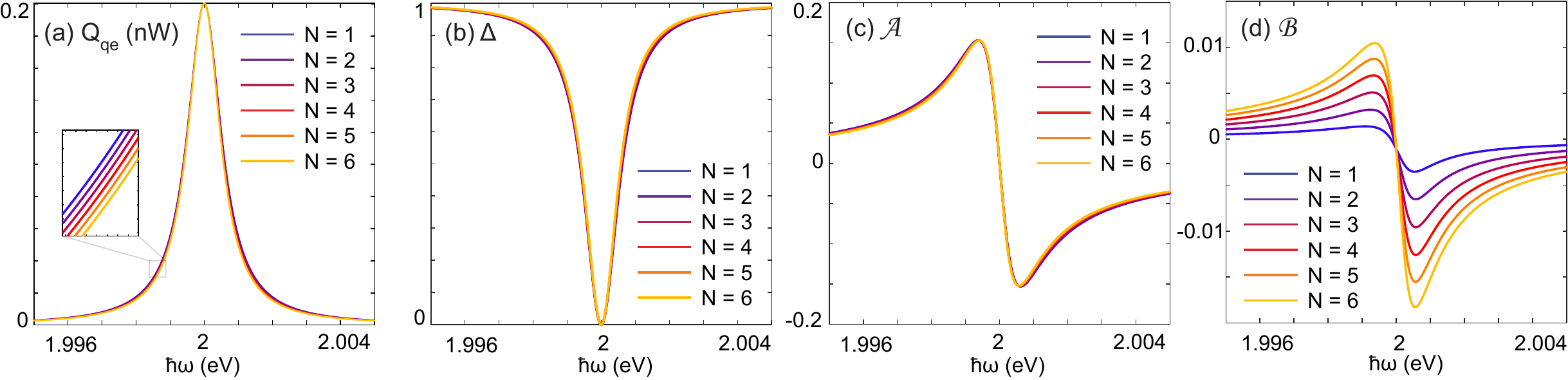}
	\centering
	\caption{(Colour online) Sample plots depicting	the variation of (a) QE absorption rate $Q_\text{qe}$, (b) QE population difference $\Delta$, (c) real and (d) imaginary parts of the slowly time varying off diagonal density matrix element component $\tilde{\rho}_\text{ge} = \mathcal{A}+i\mathcal{B}$ (where $\rho_\text{ge}=\tilde{\rho}_\text{ge} e^{i\omega t}$) under the influence of $N = 1,...,6$ MNPs, plotted against the energy $\hbar\omega$ of the coherent external illumination in a  narrow band of $\SI{5}{\milli\electronvolt}$ around the QE resonance $\hbar\omega_0$. For these sample plots, the QE resonance $\hbar\omega_0 = \SI{2}{\electronvolt}$, submerging medium permittivity $\epsilon_\text{b} = 5.45$ and the distance from QE to all MNPs, $R=\SI{15}{\nano\meter}$. \label{Fig:NarrowBand_plots}}
\end{figure*}

Throughout this section, we use the GNOR model to characterize the optical polarizability $\beta$ of the gold nanoparticles, except where we explicitly mention the use of the LRA model for comparison purposes. In Fig.\ \ref{Fig:beta_comparison}, we have illustrated the variation of GNOR based $|\beta|$ for three different (relative) permittivity values of submerging media, $\epsilon_\text{b}=1.78$, 3.3 and 5.45. It is useful to notice that the plasmonic peak amplifies and redshifts with increasing submerging medium permittivity. We have also depicted the variation of $|\beta|$ as predicted by the conventional LRA model for $\epsilon_\text{b} = 3.3$, which is seen to possess a larger, red-shifted peak compared to its nonlocal counterpart, as is also suggested by literature \cite{raza2015nonlocal}.

\subsection{The narrowband analysis}

Let us first analyse the optical response of a QE placed at the centre of an equispaced, planar MNP ring, when the external coherent illumination sweeps a narrow frequency band of $\SI{5}{\milli\electronvolt}$ around the QE resonance. Fig.\ \ref{Fig:NarrowBand_plots} depicts such sample plots we have obtained for QE absorption rate $Q_\text{qe}$, population difference $\Delta$ and the real ($\mathcal{A}$) and imaginary ($\mathcal{B}$) parts of the slowly varying off diagonal density matrix element component ($\tilde{\rho}_\text{ge} = \mathcal{A}+i\mathcal{B}$), for a system with bare excitonic resonance energy $\hbar\omega_0=\SI{2}{\electronvolt}$, submerging medium permittivity $\epsilon_\text{b} = 5.45$ and MNP-QE centre separation $R=\SI{15}{\nano\meter}$. Notice (using Fig.\ \ref{Fig:beta_comparison}) that for these sample plots, the bare excitonic resonance energy lies close to the plasmonic resonance energy of a $\SI{3}{\nano\meter}$ radius MNP at  $\epsilon_\text{b} = 5.45$ ($\hbar\omega_\text{sp}\approx\SI{2.09}{\electronvolt}$).

Subplot Fig.\ \ref{Fig:NarrowBand_plots}(b),
which depicts the variation of the QE population difference $\Delta = \rho_\text{gg} - \rho_\text{ee}$ shows that $\Delta\to1$ as the absolute detuning of the incident field frequency from the QE resonance gets larger, for all values of N. Moreover, $\Delta\to 0$ when the detuning $\to 0$ (that is, when $\omega\to\omega_0$), tracing a singly dipped spectral shape. As is also suggested by (\ref{Eq:QE_Absorption_rate}), QE absorption rate depicted in Fig.\ \ref{Fig:NarrowBand_plots}(a) follows a singly peaked spectral shape with its peak aligning with the dip of $\Delta$ at resonance, and tending to zero as the detuning increases. Also note from subplots (c) and (d) that $\mathcal{A}$ and $\mathcal{B}$ possess Fano-like lineshapes around QE resonance which tend to zero with increasing detuning, for all values of $N$ considered.

We examined a range of $Q_\text{qe}$, $\Delta$, $\mathcal{A}$ and $\mathcal{B}$ plots in a wide parameter region where $\epsilon_\text{b}$ was varied from $1.78-5.45$ and $\hbar\omega_0$ was varied from $1.5-\SI{3}{\electronvolt}$. In regions of reduced plasmonic impact on the QE, resulting from parameters/parameter combinations such as high detunings of $\omega_0$ from the plasmonic resonance ($\omega_\text{sp}$), low $\epsilon_\text{b}$ (resulting in diminishing plasmonic peaks as depicted in Fig.\ \ref{Fig:beta_comparison}), or large values of $R$, the amplitudes of both the Fano peak and trough of $\mathcal{B}$ was seen to reduce, accompanied by distortions to the Fano-shape with higher trough amplitudes in comparison to the respective peaks. In the entire parameter space examined, the singly peaked, singly dipped and Fano-like line shapes around QE resonance observed for $Q_\text{qe}$, $\Delta$ and $\mathcal{A}$, respectively, were preserved (similar to the sample plots in Fig.\ \ref{Fig:NarrowBand_plots}). However, $\SI{}{\milli\electronvolt}$ scale linewidth variations were observable for $Q_\text{qe}$ and $\Delta$, as shown in the inset in Fig.\ \ref{Fig:NarrowBand_plots}(a). We analyse these linewidth variations in the next section.

\subsection{Analysis of linewidth variations}

\begin{figure*}[t!]
	\includegraphics[width=\textwidth]{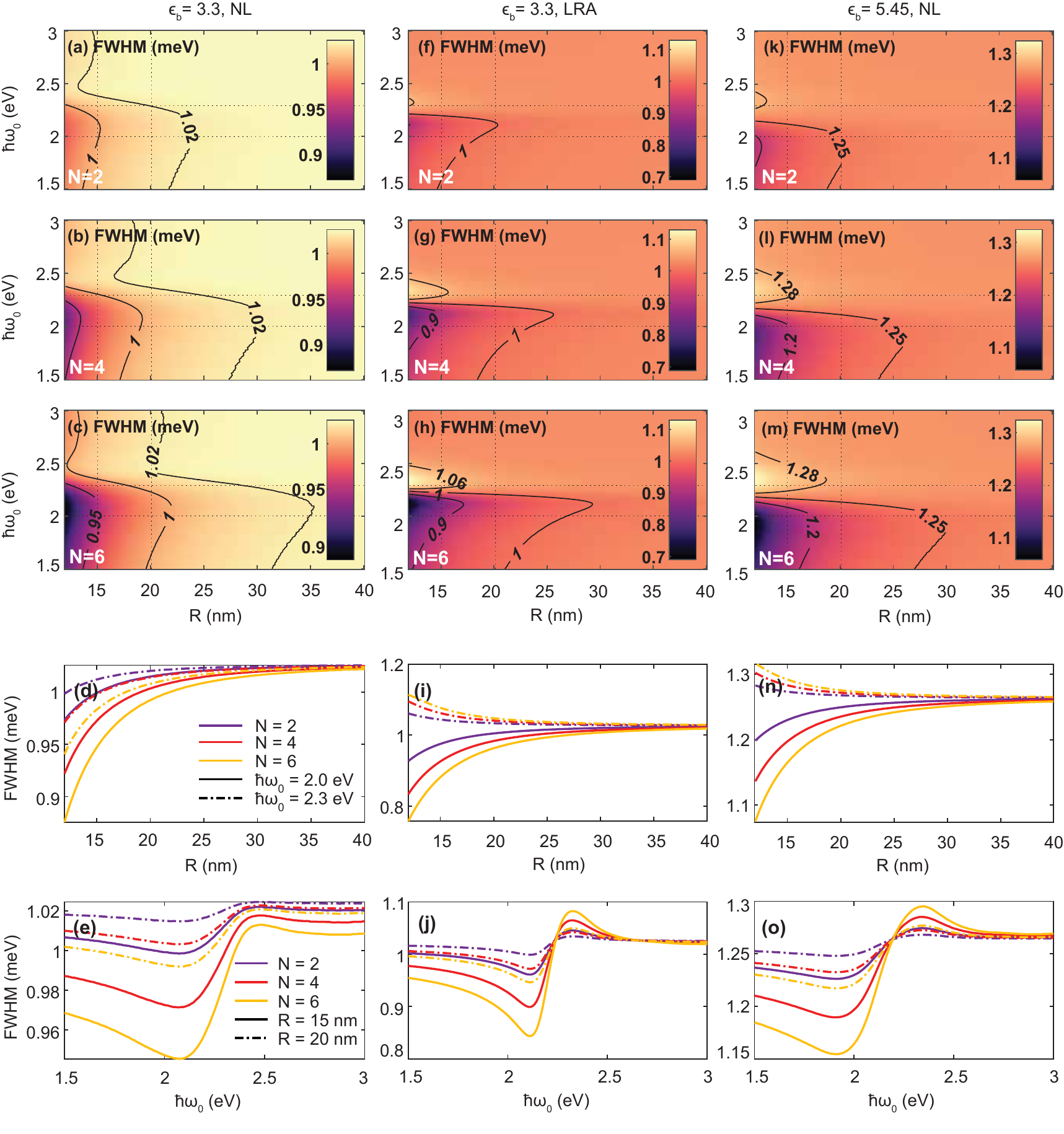}
	\centering
	\caption{(Colour online) Variation of full width at half maximum (FWHM) of the QE absorption rate ($Q_\text{qe}$) in the presence of $N$ equispaced spherical MNPs. Subplots (a), (b) and (c) of the first column depict the top-view of the FWHM surface plots in the presence of $N=2$, 4 and 6 MNPs respectively, plotted against the centre separation of each MNP from the QE ($R$) and the bare excitonic resonance energy ($\hbar\omega_0$), when the system is submerged in a medium of permittivity $\epsilon_\text{b} = 3.3$ and the polarizability of each MNP is modelled nonlocally (NL), using the GNOR formalism. Subplot (d) depicts the 2D view of the horizontal cross sections (denoted as black dotted lines at $\hbar\omega_0=\SI{2}{\electronvolt}$ and $\SI{2.3}{\electronvolt}$) in the three preceding FWHM surface plots, where solid lines of respective colour represent the cross sections at $\hbar\omega_0 = \SI{2}{\electronvolt}$ and the dotted-dashed lines represent those at  $\hbar\omega=\SI{2.3}{\electronvolt}$, for $N=2$, 4 and 6. Similarly, subplot (e) depicts the vertical cross sections shown as dashed lines in the FWHM surface plots (a), (b) and (c), where the solid lines represent the cross sections at $R=\SI{15}{\nano\meter}$ and the dotted-dashed lines represent those at $R=\SI{20}{\nano\meter}$. The second and third columns represent similar arrangements of plots for cases where $\epsilon_\text{b}=3.3$ with the MNP polarizability modelled using the local response approximation (LRA), and for $\epsilon_\text{b}=5.45$ modelled using the nonlocal  GNOR formalism, respectively. The legends shown in subplots (d) and (e) are common to all plots in the row. \label{Fig:FWHM_surf}} 
\end{figure*}

We analysed, using both nonlocal GNOR (NL) and local response approximation (LRA) based models, the variation of the full width at half maximum (FWHM) of $Q_\text{qe}$ against the bare excitonic resonance energy ($\hbar\omega_0$) and MNP-QE centre separation ($R$) for a number of cases, where the submerging medium permittivity $\epsilon_\text{b}$ was varied from $1.78-5.45$. For all cases within the aforementioned parameter range, the convergence factor lay below 0.5, suggesting feedback convergence and hence the safe usability of our model. Sample results obtained in our analysis for $\epsilon_\text{b}=3.3$ (using both NL and LRA formalisms) and $\epsilon_\text{b} = 5.45$ (using NL formalism), are presented in Fig.\ \ref{Fig:FWHM_surf}. 

\begin{figure*}[t!]
	\includegraphics[width=\textwidth]{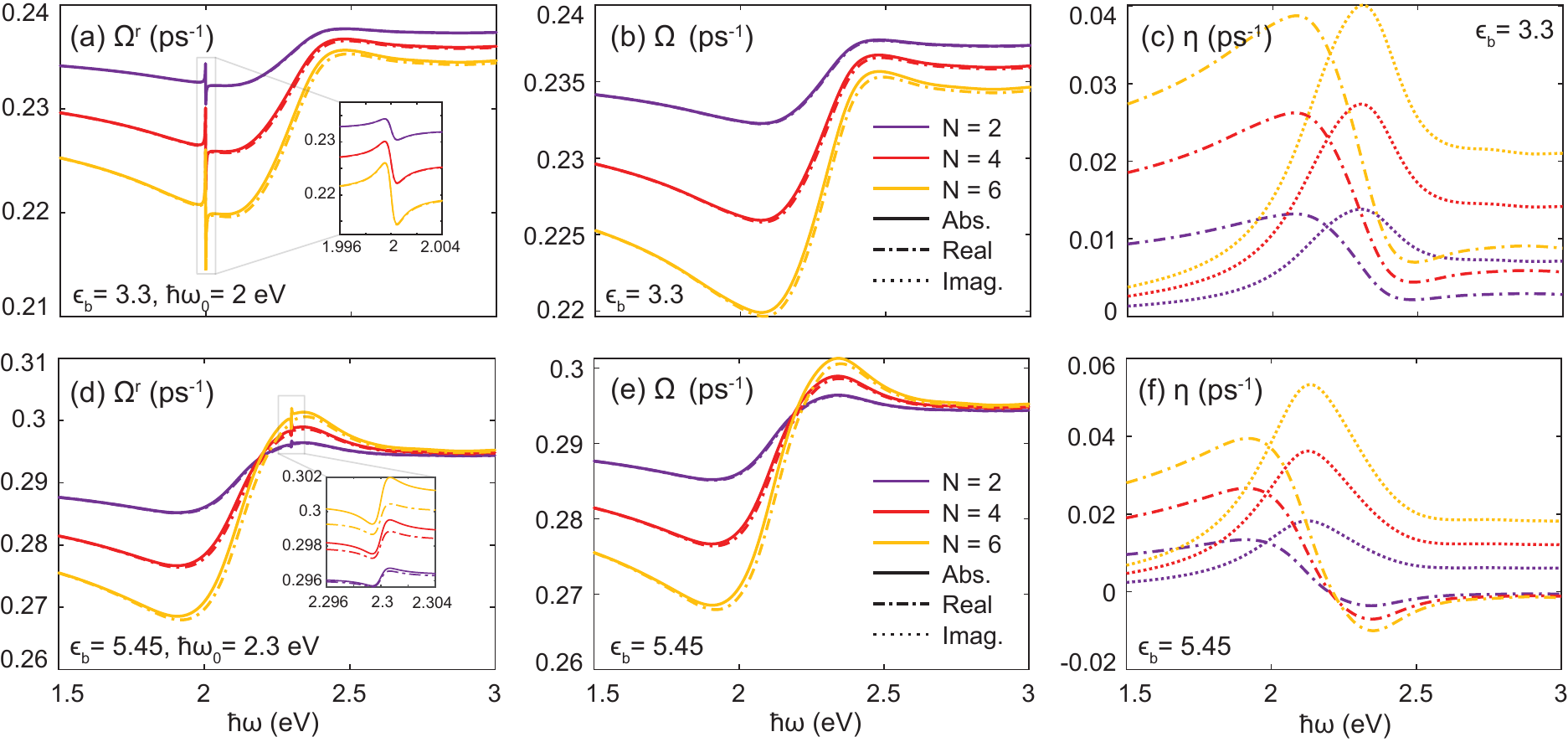}
	\centering
	\caption{(Colour online) (a) Effective or normalized Rabi frequency $\Omega^r$ experienced by a quantum emitter with bare excitonic energy $\hbar\omega_0=\SI{2}{\electronvolt}$ under the influence of $N=2$, 4 and 6 equispaced MNPs, each located at a distance $R=\SI{15}{\nano\meter}$ from the QE, when the entire system is submerged in a medium of relative permittivity $\epsilon_\text{b} = 3.3$, (b) The corresponding Rabi frequency in the absence of coherences ($\Omega$), (c) The coefficient of QE self-interaction ($\eta$). The legend in the middle subplot is common to the entire row where the solid, dotted-dashed and dotted lines correspond to the relevant absolute, real and imaginary quantities, respectively. The second row depicts an arrangement of plots similar to the earlier, for the parameters $\epsilon_\text{b} = 5.45$, $\hbar\omega_0 = \SI{2.3}{\electronvolt}$ and $R=\SI{15}{\nano\meter}.$ Insets in subplots (a) and (d) are enlargements along the frequency axis around $\hbar\omega_0$, for better visibility of the spectral signatures.\label{Fig:Rabi_Om_eta}}
\end{figure*}

Observation of all surface plots depicted in Fig.\ \ref{Fig:FWHM_surf} reveals the existence of linewidth variation hotspots (in comparison to the respective far field linewidths) near the regions where the bare excitonic resonance lies close to the relevant plasmonic resonance ($\omega_0\approx\omega_\text{sp}$). Moreover, it can be seen that these linewidth variations amplify with decreasing $R$ and increasing $N$ which clearly implies contribution of the plasmonically induced fields. We also observe that for some cases, (for example, the nonlocally modelled $\epsilon_\text{b}=3.3$ case in Fig.\ \ref{Fig:FWHM_surf}) the linewidth decreases with decreasing MNP-QE separation ($R$) for all values of $\omega_0$ under study, whereas for other cases (generally, with larger $\max{\abs{\beta}}$ values), both increasing and decreasing of linewidth against decreasing $R$ were observable in different regions of $\omega_0$. To analyse these variations further and gain more insight into the plasmonic contributions, we studied both horizontal and vertical cross sections of the FWHM surface plots, a few samples of which are shown in the fourth and fifth rows of Fig.\ \ref{Fig:FWHM_surf}.

As it is observable from the horizontal surface plot cross sections in the the fourth row of Fig.\ \ref{Fig:FWHM_surf}, linewidth variation of the $Q_\text{qe}$ spectrum (in comparison to the respective far field value) increases with the number of MNPs $N$ for a given value of $R$ and $\omega_0$. That is, irrespective of whether FWHM shows an increasing or a decreasing trend against decreasing $R$, curves with larger values of $N$ show larger linewidth variations (observe how $N=6$ curves are the outermost and $N=2$ curves are the inner most in all three subplots (d), (i) and (n) of Fig.\ \ref{Fig:FWHM_surf}). This is a clear indication of the observed linewidth variation phenomenon being driven by the resultant plasmonic field experienced by the QE.

When we shifted our attention towards the vertical FWHM surface plot cross sections (a few samples of which are depicted in the last row of Fig.\ \ref{Fig:FWHM_surf}), we observed a striking resemblance between the shapes of FWHM vs $\hbar\omega_0$ plots and the respective $\Omega$ vs $\hbar\omega$ plots. This observation was consistent across the entire parameter range we studied. Compare the shape of Fig.\ \ref{Fig:FWHM_surf}(e) against Fig.\ \ref{Fig:Rabi_Om_eta}(b) and Fig.\ \ref{Fig:FWHM_surf}(o) against Fig.\ \ref{Fig:Rabi_Om_eta}(e) to observe this resemblance. It is evident that this observation further validates our earlier claims of plasmonic-field influencing the QE absorption linewidth variations. Furthermore, we observed that the increase or decrease of FWHM with decreasing $R$ (see the sample plots in the fourth row of Fig.\ \ref{Fig:FWHM_surf}) was associated with the narrow Fano lineshape observed near $\omega\approx\omega_0$ in the respective Rabi frequency ($\Omega^r$) spectra, throughout the parameter region studied. Let us analyse this association using the sample spectra of $\Omega^r$, $\Omega$ and $\eta$ in Fig.\ \ref{Fig:Rabi_Om_eta}.


\begin{figure*}[t!]
	\includegraphics[width=\textwidth]{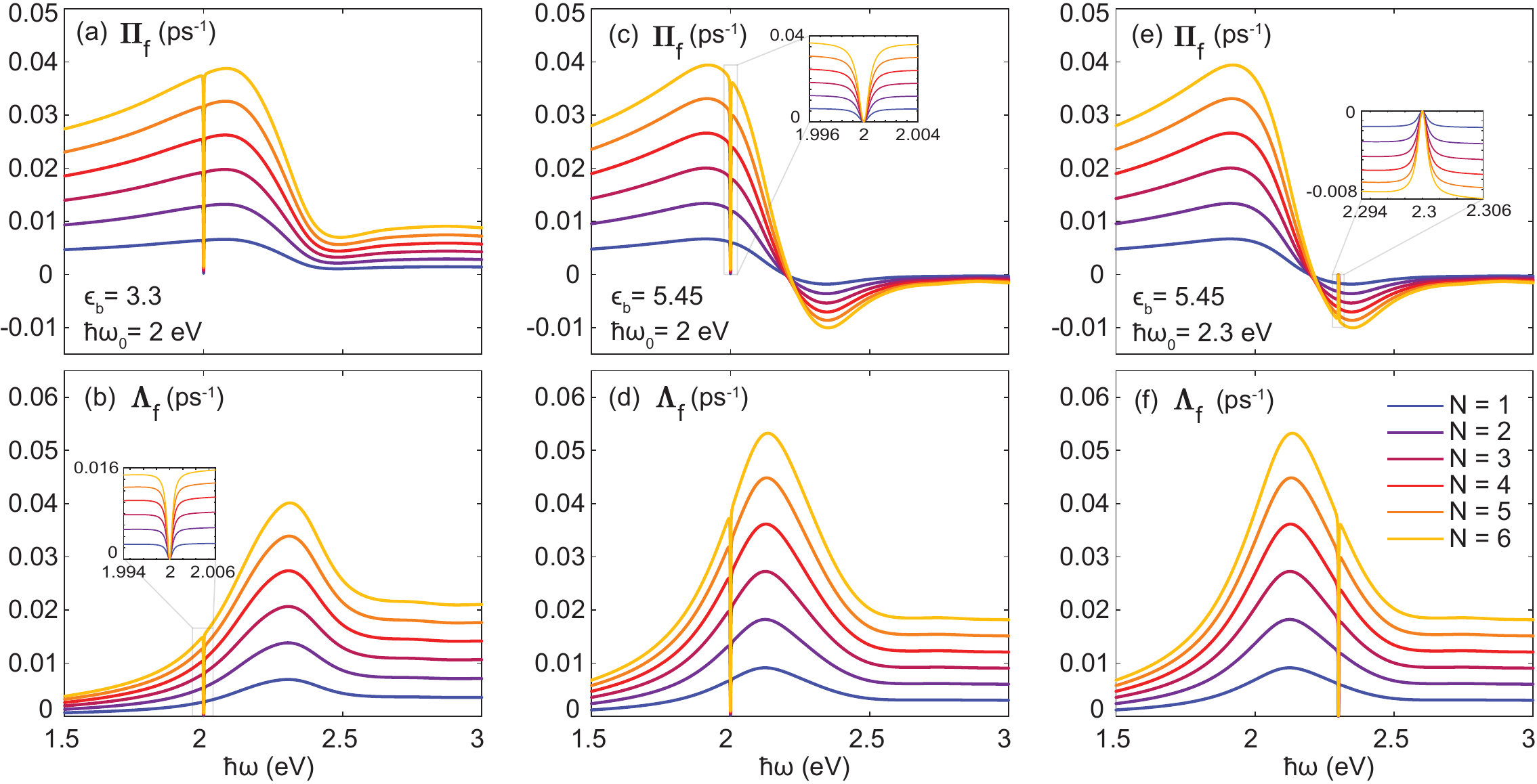}
	\centering
	\caption{(Colour online) (a) The exciton transition energy redshift ($\Pi_f$) experienced by a quantum emitter with bare excitonic energy $\hbar\omega_0 = \SI{2}{\electronvolt}$ surrounded by $N=1$,2,...6 MNPs, each located at a distance $R=\SI{15}{\nano\meter}$ and submerged in a medium of relative permittivity $\epsilon_\text{b}=3.3$ (b) Exciton dephasing rate blueshift ($\Lambda_f$) experienced by the same QE. The second and third columns show similar constellations of plots for $\epsilon_\text{b} = 5.45$, $\hbar\omega_0 = \SI{2}{\electronvolt}$ and $\epsilon_\text{b}=5.45$, $\hbar\omega_0=\SI{2.3}{\electronvolt}$, respectively. The insets in subplots (b), (c) and (e) are enlargements along the frequency axis around $\hbar\omega_0$, for better visibility of the spectral signatures. \label{Fig:Pi_f_lambda_f}}
\end{figure*}

The first row of Fig.\ \ref{Fig:Rabi_Om_eta} depicts the spectra for $\Omega^r$ (when $\hbar\omega_0 = \SI{2}{\electronvolt}$), $\Omega$ and $\eta$ for the (nonlocally modelled) case where $\epsilon_\text{b}=3.3$. We can readily observe that the absolute values of both $\Omega^r$ and $\Omega$ closely align with the respective real valued components, suggesting a minimal contribution from the respective imaginary parts. Notice again, how the shape of $\Re{\Omega}$ spectrum strikingly resembles the shape of the FWHM vs $\hbar\omega_0$ curve for $\epsilon_\text{b} = 3.3$ in Fig.\ \ref{Fig:FWHM_surf}(e). Let us now recall from our formalism section that $\Omega^r = \Omega + \eta(\mathcal{A}-i\mathcal{B})$, extraction of the real part of which yields, $\Re{\Omega^r} = \Re{\Omega}+\Re{\eta}\mathcal{A}+\Im{\eta}\mathcal{B}$. As $\Omega$ is independent of QE contributions, the narrow Fano-shaped signature of $\Omega^r$ near $\omega\approx\omega_0$ forms mainly due to the contributions from $\Re{\eta}\mathcal{A}+\Im{\eta}\mathcal{B}$ when $\abs{\Omega^r}\approx\Re{\Omega^r}$, where $\mathcal{A}$ and $\mathcal{B}$ take the narrow Fano-like lineshapes around QE resonance we observed in Fig.\ \ref{Fig:NarrowBand_plots}.

If we vary $\hbar\omega_0$ in Fig.\ \ref{Fig:Rabi_Om_eta}(a) from $1.5-\SI{3}{\electronvolt}$, the narrow Fano-like lineshape now observed around $\hbar\omega\approx\SI{2}{\electronvolt}$ will traverse from $\approx 1.5 - \SI{3}{\electronvolt}$ without a reversal of shape. That is, the Fano-like lineshape we observed for $\mathcal{A}$ and $\mathcal{B}$, where a peak is followed by a trough, will be qualitatively retained in $\Re{\Omega^r}$'s Fano signature around QE resonance. However, as $\hbar\omega_0$ traverses from $1.5-\SI{3}{\electronvolt}$, the amplitude variation of the Fano peak and trough will be governed by $\Re{\eta}$ and $\Im{\eta}$ indicated in Fig.\ \ref{Fig:Rabi_Om_eta}(c). When both $\Re{\eta}$ and $\Im{\eta}$ take large positive values, the similarly shaped Fano patterns of $\mathcal{A}$ and $\mathcal{B}$ will be linearly (additively) combined to give the enhanced Fano pattern we observe for $\Re{\Omega^r} = \Re{\Omega} + \Re{\eta}\mathcal{A} + \Im{\eta}\mathcal{B}$ around $\omega\approx\omega_0$ when $\hbar\omega_0=\SI{2}{\electronvolt}$ in Fig.\ \ref{Fig:Rabi_Om_eta}(a). When $\hbar\omega_0\to\SI{1.5}{\electronvolt}$ or $\SI{3}{\electronvolt}$ (the two extreme points along the x-axis), the peak and trough amplitudes of $\Re{\Omega^r}$'s narrow Fano pattern will comparatively reduce due to the diminishing but positive $\Re{\eta}$ and $\Im{\eta}$ values observed in Fig.\ \ref{Fig:Rabi_Om_eta}(c) at the two extreme ends along the x-axis. Notice how the Fano signature of $\Re{\Omega^r}$ enhances with increasing $N$ due to increased plasmonic impact, which is also explainable by the aforementioned formula for $\Re{\Omega^r}$. In essence, when both $\Re{\eta}$ and $\Im{\eta}$ are positive, the Fano signature of $\Re{\Omega^r}$ will not reverse in shape along the $x-$axis, but its magnitude will be governed by $\Re{\eta}$ and $\Im{\eta}$.  

If we now shift our attention to Fig.\ \ref{Fig:Rabi_Om_eta}(d) which depicts the variation of real and absolute spectra of $\Omega^r$ at $\epsilon_\text{b} = 5.45$ and $\hbar\omega_0 = \SI{2.3}{\electronvolt}$, we can observe a reversed and diminished Fano signature around $\hbar\omega_0$ in comparison to our earlier observation in Fig.\ \ref{Fig:Rabi_Om_eta}(a). This variation too can be explained using our earlier formula $\Re{\Omega^r} = \Re{\Omega} + \Re{\eta}\mathcal{A} + \Im{\eta}\mathcal{B}$ as follows. Comparison of Fig.\ \ref{Fig:Rabi_Om_eta}(d) against Fig.\ \ref{Fig:Rabi_Om_eta}(f) reveals that $\Re{\eta}$ is now negative near $\hbar\omega_0=\SI{2.3}{\electronvolt}$. Such negation of $\Re{\eta}$ is highly likely to reverse the Fano lineshape resulting from the $\Re{\eta}\mathcal{A} + \Im{\eta}\mathcal{B}$ component of $\Re{\Omega^r}$ as $\mathcal{A}$ is usually sufficiently larger in magnitude than $\mathcal{B}$ (for example, see Fig.\ \ref{Fig:NarrowBand_plots}). The diminishing of Fano amplitude results from the competition between the now differently signed $\mathcal{A}$ and $\mathcal{B}$. 

For all our observations in the wide parameter region considered (for both LRA and GNOR based models), the MNP-QE constellations exhibiting patterns of increasing FWHM against decreasing $R$ (for example, the dotted dashed lines in Fig.\ \ref{Fig:FWHM_surf}(i) and (n)) could be mapped to regions of negative $\Re{\eta}$, and hence reversed Fano-shapes of $\Re{\Omega^r}$ near the respective excitonic resonance $\hbar\omega_0$, as explained above. On the contrary, the cases exhibiting patterns of decreasing FWHM against decreasing $R$ (for example, all curves in Fig.\ \ref{Fig:FWHM_surf}(d) and the solid lines in Fig.\ \ref{Fig:FWHM_surf}(i) and (n)) could be mapped to regions of positive $\Re{\eta}$, and hence non-reversed Fano-shapes of $\Re{\Omega^r}$ near $\hbar\omega_0$.

\subsection{Coupled-plasmon induced excitonic energy and dephasing rate shifts}

Let us finally analyse the influence of the weakly inter-coupled plasmonic ring of equispaced spherical MNPs on the excitonic energy and the dephasing rate of the QE at the centre. In the first and second rows of Fig.\ \ref{Fig:Pi_f_lambda_f}, we have shown the spectra of QE exciton transition energy redshift ($\Pi_f=\Re{\eta}\Delta$) and QE dephasing rate blueshift ($\Lambda_f=\Im{\eta}\Delta$) due to the surrounding $N=1,...,6$ MNPs, for three sample cases where $\epsilon_\text{b}=3.3$, $\hbar\omega_0=\SI{2}{\electronvolt}$ (first column), $\epsilon_\text{b}=5.45$, $\hbar\omega_0=\SI{2}{\electronvolt}$ (second column) and $\epsilon_\text{b}=5.45$ and $\hbar\omega_0=\SI{2.3}{\electronvolt}$ (third column). As is also suggested by their equations, $\Pi_f$ and $\Lambda_f$ exactly follow the variations of $\Re{\eta}$ and $\Im{\eta}$, respectively, except at the sharp narrow slits towards zero around QE resonance, resulting from the spectrum of $\Delta$ (which goes to zero near QE resonance and to 1 as the external field detunes from the QE resonance, as we saw in Fig.\ \ref{Fig:NarrowBand_plots}). As $\Im{\eta}>0$ for all three cases observed, the QE experiences dephasing rate blueshifts in the entire range where $\hbar\omega$ varies from $\SI{1.5}{\electronvolt}$ to $\SI{3}{\electronvolt}$, except when $\omega=\omega_0$ where $\Lambda_f=0$, for all values of $N$. The observed dephasing rate blueshift is seen to increase with $N$, throughout the spectral region studied. 

It is evident that the QE experiences plasmon-induced transition energy redshifts in the regions where $\Re{\eta}>0$ and blueshifts where $\Re{\eta}<0$ (except at $\omega=\omega_0$). Thus, our results suggest that, for QEs under the influence of weakly inter-coupled symmetric planar MNP systems, parameter regions exhibiting trends of decreasing $Q_\text{qe}$ linewidths against decreasing MNP-QE centre separations $R$ are likely to be associated with plasmon induced excitonic energy redshifts (for example, compare Fig.\ \ref{Fig:FWHM_surf}(d) with Fig.\ \ref{Fig:Pi_f_lambda_f}(a)), as both these phenomena occur when $\Re{\eta}>0$. Similarly, regions exhibiting trends of increasing $Q_\text{qe}$ linewidths against decreasing MNP-QE centre separations $R$ are likely to be associated with plasmon induced excitonic energy blueshifts (compare the dotted-dashed lines in Fig.\ \ref{Fig:FWHM_surf}(n) with Fig.\ \ref{Fig:Pi_f_lambda_f}(e)), as both these phenomena occur when $\Re{\eta}<0$. Moreover, irrespective of the shift type (red/blueshift), the magnitude of the plasmon induced excitonic energy shift tends to increase with $N$. 

\section{Conclusion}
We developed an analytical framework to study the influence of a weakly intercoupled inplane spherical metal nanoparticle (MNP) assembly on a coherently illuminated quantum emitter (QE), using the generalized nonlocal optical response (GNOR) theory. We reduced the derived equations into simple and concise expressions representing a QE mediated by a symmetric MNP setup. Using the derived model, we analysed the optical properties of a coherently illuminated QE at the centre of an inplane symmetric MNP setup. We observed that the QE experiences plasmon induced absorption rate spectral linewidth variations that increase in magnitude with decreasing MNP-QE centre separation and increasing number of MNPs. We could also observe that the parameter regions where the QE exhibits trends of decreasing linewidth against decreasing MNP-QE centre separation are likely to be associated with plasmon induced excitonic energy redshifts, and vice versa. The magnitude of the observed exciton energy red/blueshift was seen to increase with the number of MNPs in the constellation.\\

\section*{Acknowledgements}
HH gratefully acknowledges the encouragement and support by D. U. Kudavithana. This research was funded by the Australian Research Council under grant number CE170100026.

\renewcommand{\theequation}{A-\arabic{equation}}
  \setcounter{equation}{0}  
\section*{Appendix}\label{Sec:Appendix}
\subsection*{Obtaining the external field feedback dipoles}

The $0^\text{th}$ level external field feedback dipole, or the dipole moment component formed directly due to the external field (with a positive frequency component $E^+ = E_0 e^{-i\omega t}$), in an $f_0^\text{th}$ MNP can be found using (\ref{Eq:Induced_dipole}) as,
\begin{equation}\label{Eq:Induced_dipole_0th}
d^+_{f_0} = (4\pi\epsilon_0\epsilon_\text{b}) r_{f_0}^3\beta_{f_0}E_0e^{-i\omega t}.
\end{equation}
Collective first level feedback dipole induced in the $\text{n}^\text{th}$ MNP due to all such $0^\text{th}$ level dipoles (where the dipole index $f_0\neq \text{n}$) can be obtained with the aid of (\ref{Eq:Dipole_induced_dipole}) as,
\begin{align}\label{Eq:First_level_feedback_non_sym}
	d_{\text{n}\_\Sigma f_0}^+ &= (4\pi\epsilon_0\epsilon_\text{b})r_\text{n}^3\beta_\text{n}s_\alpha E_0 e^{-i\omega t}\sum_{\substack{f_0=1 \\ f_1 \neq \text{n}}}^N\left(\frac{r_{f_0}^3\beta_{f_0}}{R_{\text{n}{f_0}}^3}\right)\nonumber\\
	&=d_\text{n}^+ s_\alpha \sum_{\substack{f_0=1 \\ f_0 \neq \text{n}}}^N\left(\frac{r_{f_0}^3\beta_{f_0}}{R_{\text{n}{f_0}}^3}\right).
\end{align}
Similarly, using the first level feedback pattern above, the second level external field feedback dipole induced in the $\text{n}^\text{th}$ MNP due to the collective first level external field feedback dipoles induced in all surrounding MNPs (indexed in the summation using the subscript $f_1$, where $f_1\neq n$) can be obtained as,
\begin{align}
	d_{\text{n}\_\Sigma f_1\_\Sigma f_0}^+ = d_\text{n}^+ s_\alpha^2 \sum_{\substack{f_1=1 \\ f_1 \neq \text{n}}}^N\left(\frac{r_{f_1}^3\beta_{f_1}}{R_{\text{n}{f_1}}^3} \sum_{\substack{f_0=1 \\ f_0 \neq f_1}}^N \frac{r_{f_0}^3\beta_{f_0}}{R_{{f_1}{f_0}}^3}\right).
\end{align}
By repeating the above procedure, the $\text{p}^\text{th}$ level feedback dipole formed in the $\text{n}^\text{th}$ MNP due to the collective $(\text{p}-1)^\text{th}$ level dipoles in the surrounding MNPs (indexed in the summation using the subscript $f_{\text{p}-1}$) can be found, which results in equation (\ref{Eq:P_th_level_dipole}) in the main text.

\providecommand{\noopsort}[1]{}\providecommand{\singleletter}[1]{#1}%
%

\end{document}